\documentclass[12pt]{article}
\usepackage{amsmath}  
\input amssym

\begin{document}

\def\prg#1{\medskip\noindent{\bf #1}}     \def\ra{\rightarrow}
\def\lra{\leftrightarrow}        \def\Ra{\Rightarrow}
\def\nin{\noindent}              \def\pd{\partial}
\def\dis{\displaystyle}          \def\dfrac{\dis\frac}
\def\grl{{GR$_\Lambda$}}         \def\vsm{\vspace{-9pt}}
\def\Lra{{\Leftrightarrow}}      \def\ads3{AdS$_3$}
\def\ads3{{\rm AdS$_3$}}
\def\Leff{\hbox{$\mit\L_{\hspace{.6pt}\rm eff}\,$}}
\def\bull{\raise.25ex\hbox{\vrule height.8ex width.8ex}}
\def\Tr{\hbox{\rm Tr\hspace{1pt}}}
\def\as{{\rm as}}     \def\inn{\,\rfloor\,}
\def\ric{{(Ric)}}     \def\tmgl{\hbox{TMG$_\Lambda$}}
\def\nmg{{NMG}}       \def\nmgl{\hbox{NMG$_\Lambda$}}
\def\mb#1{\mbox{\boldmath{$#1$}}}
\def\hd{\hspace{2pt}{}^\star\hspace{-1pt}}

\def\cL{{\cal L}}     \def\cM{{\cal M }}    \def\cE{{\cal E}}
\def\cH{{\cal H}}     \def\hcH{\hat{\cH}}   \def\bcH{{\bar\cH}}
\def\cK{{\cal K}}     \def\hcK{\hat{\cK}}   \def\bcK{{\bar\cK}}
\def\cO{{\cal O}}     \def\hcO{\hat{\cO}}   \def\tR{{\tilde R}}
\def\cB{{\cal B}}     \def\bV{{\bar V}}     \def\heps{\hat\epsilon}
\def\cT{{\cal T}}     \def\D{{\Delta}}      \def\L{{\mit\Lambda}}
\def\hcO{{\hat\cO}}   \def\hcH{\hat\cH}     \def\bL{{\L_0}}
\def\bu{{\bar u}}     \def\bv{{\bar v}}     \def\bw{{\bar w}}
\def\cR{{\cal R}}     \def\bcR{{\bar\cR}}   \def\bz{{\bar z}}
\def\hcR{{\hat\cR}}   \def\barG{{\bar G}}   \def\barf{{\bar f}}
\def\barb{{\bar b}}   \def\barl{{\bar\l}}   \def\bQ{{\bar Q}}
\def\tQ{{\tilde Q}}

\def\G{\Gamma}        \def\S{\Sigma}
\def\a{\alpha}        \def\b{\beta}          \def\g{\gamma}
\def\d{\delta}        \def\m{\mu}            \def\n{\nu}
\def\th{\theta}       \def\k{\kappa}         \def\l{\lambda}
\def\vphi{\varphi}    \def\ve{\varepsilon}   \def\p{\pi}
\def\r{\rho}          \def\Om{\Omega}        \def\om{\omega}
\def\s{\sigma}        \def\t{\tau}           \def\eps{\epsilon}
\def\nab{\nabla}      \def\Ups{{\mit\Upsilon}}
\def\Th{\Theta}       \def\cT{{\cal T}}      \def\cS{{\cal S}}
\def\cV{{\cal V}}     \def\hR{{\hat R}{}}
\def\f{{f}}           \def\tom{{\tilde\om}}
\def\tb{{\tilde b}}   \def\tf{{\tilde f}}    \def\tl{{\tilde\lambda}}
\def\tpi{{\tilde\pi}} \def\tPi{{\tilde\Pi}}  \def\tH{\tilde H}
\def\tp{{\tilde p}}   \def\tP{{\wt P}}       \def\tcL{{\tilde\cL}}
\def\wt{\widetilde}   \def\ol{\overline}
\def\tcT{{\wt\cT}}
\def\tcK{{\wt\cK}}    \def\tcH{{\wt\cH}}    \def\tcR{{\wt\cR}}
\def\tG{\tilde G}     \def\tu{{\tilde u}}   \def\tv{{\tilde v}}
\def\tw{{\tilde w}}   \def\tz{{\tilde z}}   \def\tvphi{{\tilde\vphi}}
\def\tphi{{\tilde\phi}} \def\tPhi{{\wt\Phi}}
\def\nn{\nonumber}
\def\be{\begin{equation}}             \def\ee{\end{equation}}
\def\ba#1{\begin{array}{#1}}          \def\ea{\end{array}}
\def\bea{\begin{eqnarray} }           \def\eea{\end{eqnarray} }
\def\beann{\begin{eqnarray*} }        \def\eeann{\end{eqnarray*} }
\def\beal{\begin{eqalign}}            \def\eeal{\end{eqalign}}
\def\lab#1{\label{eq:#1}}             \def\eq#1{(\ref{eq:#1})}
\def\bsubeq{\begin{subequations}}     \def\esubeq{\end{subequations}}
\def\bitem{\begin{itemize}}           \def\eitem{\end{itemize}}
\renewcommand{\theequation}{\thesection.\arabic{equation}}

\title{Extra gauge symmetries 
       in BHT gravity}

\author{M. Blagojevi\'c and B. Cvetkovi\'c\footnote{
        Email addresses: {\tt mb@ipb.ac.rs,
                               cbranislav@ipb.ac.rs}} \\
University of Belgrade, Institute of Physics,\\
P. O. Box 57, 11001 Belgrade, Serbia}
\date{}
\maketitle

\begin{abstract}
We study the canonical structure of the Bergshoeff-Hohm-Townsend
massive gravity, linearized around a maximally symmetric background. At
the critical point in the space of parameters, defined by
$\L_0/m^2=-1$, we discover an extra gauge symmetry, which reflects the
existence of the partially massless mode. The number of the Lagrangian
degrees of freedom is found to be 1. We show that the canonical
structure of the theory at the critical point is unstable under
linearization.
\end{abstract}

\section{Introduction}
\setcounter{equation}{0}

Recently, Bergshoeff, Hohm and Townsend (BHT) proposed a parity
conserving theory of gravity in three dimensions (3D), which is defined
by adding certain curvature-squared terms to the Einstein-Hilbert
action \cite{1,2}. When the BHT gravity is linearized around the
Minkowski ground state, it is found to be equivalent to the Fierz-Pauli
theory for a free massive spin-2 field \cite{3}. Moreover, it is
ghosts-free, unitary and renormalizable \cite{4,5}. On the other hand,
the overall picture is changed when we go over to the (A)dS background,
where various dynamical properties, such as unitarity, gauge invariance
or boundary behavior, become more complex \cite{2,6,7,8,9}.

Dynamical characteristics of a gravitational theory take a particularly
clear form in the constrained Hamiltonian approach \cite{10}. Analyzing
the nature of constraints in the fully \emph{nonlinear} BHT gravity, we
discovered the special role of an extra condition \cite{11}; when
applied to a maximally symmetric solution, it takes the familiar form
$\L_0/m^2\ne -1$, where $m^2$ is the mass parameter and $\L_0$ a
cosmological constant\footnote{The canonical analysis of the BHT
gravity performed in \cite{12} refers to the case $\L_0=0$.}. The
resulting theory is found to possess two Lagrangian degrees of freedom,
in agreement with the number of massive graviton modes on the (A)dS
background \cite{2}. In the present paper, we extend our investigation
to the \emph{critical point} $\L_0/m^2=-1$ in the maximally symmetric
sector of the theory; in this case, the ground state is uniquely
determined by an effective cosmological constant \cite{2,6}. In the
linear approximation, there appears an \emph{extra gauge invariance}
which eliminates one component of the massive graviton, reducing it to
the partially massless mode \cite{13,14,15}. By comparing these results
with those obtained nonperturbatively \cite{11}, we can understand how
the canonical structure of the BHT gravity is changed in the process of
linearization. In Ref. \cite{16}, the canonical analysis of the
linearized BHT gravity is carried out only for the generic values of
the parameters.

The paper is organized as follows. In section 2, we give an account of
the linearized BHT gravity in the Lagrangian formalism. In particular,
we discuss the Lagrangian form of the extra gauge symmetry, constructed
later by the canonical methods. In section 3, we perform a complete
canonical analysis of the linearized BHT gravity around a maximally
symmetric background, assuming the critical condition $\L_0/m^2=-1$.
Then, in section 4, we classify the constraints and find a difference
in their number and type (first or second class), in comparison to the
results of the nonperturbative analysis \cite{11}. As a consequence, we
conclude that the theory exhibits a single Lagrangian degree of
freedom. In section 5, the resulting set of constraints is used to
construct the canonical generator of extra gauge symmetry. After that,
the existing Lagrangian mode can be interpreted as a partially massless
state. Finally, section 6 is devoted to concluding remarks, while
appendices contain some technical details.

We use the same conventions as in Ref. \cite{11}: the Latin indices
refer to the local Lorentz frame, the Greek indices refer to the
coordinate frame;  the middle alphabet letters
$(i,j,k,...;\m,\n,\l,...)$ run over 0,1,2, the first letters of the
Greek alphabet $(\a,\b,\g,...)$ run over 1,2; the metric components in
the local Lorentz frame are $\eta_{ij}=(+,-,-)$; totally antisymmetric
tensor $\ve^{ijk}$ and the tensor density $\ve^{\m\n\r}$ are both
normalized by $\ve^{012}=1$.

\section{Linearized Lagrangian dynamics}
\setcounter{equation}{0}

Following the approach defined in our previous paper \cite{11}, we
study the BHT gravity in the framework of Poincar\'e gauge theory
\cite{17}, where the basic gravitational variables are the triad field
$b^i$ and the Lorentz connection $\om^k$ (1-forms), and the
corresponding field strengths are the torsion
$T^i=db^i+\ve^i{}_{jk}\om^j\wedge b^k$ and the curvature
$R^i=d\om^i+\frac{1}{2}\,\ve^i{}_{jk}\om^j\wedge\om^k$ (2-forms). The
underlying geometric structure corresponds to Riemann-Cartan geometry,
in which $b^i$ is an orthonormal coframe, $g:=\eta_{ij}b^i\otimes b^j$
is the metric of spacetime, and $\om^i$ is the Cartan connection. For
$T_i=0$, the geometry becomes Riemannian.

\prg{Lagrangian.} In local coordinates $x^\m$, the BHT Lagrangian
density can be written in the form \cite{11}:
\bsubeq\lab{2.1}
\be
\cL=a\ve^{\m\n\r}\left(\s b^i{_\m}R_{i\n\r}
    -\frac{1}{3}\bL\ve^{ijk}b_{i\m}b_{j\n}b_{\k\r}\right)
    +\frac{a}{m^2}\cL_K+\ve^{\m\n\r}\frac{1}{2}\l^i{_\m}T_{i\n\r}\, .
\ee
Here, the Lagrange multiplier $\l^i{_\m}$ ensures the vanishing of
torsion and thereby, the Riemannian nature of the connection, while
$\cL_K$ is defined in terms of an auxiliary field $f^i{_\m}$ as
\be
\cL_K=\frac{1}{2}\ve^{\m\n\r}\f^i{_\m}R_{i\n\r}-b\cV_K\,,\qquad
\cV_K=\frac{1}{4}\left(f_{i\m}f^{i\m}-f^2\right)\, ,
\ee
\esubeq
where $f:=f^k{_\r}h_k{^\r}$ and $b=\det(b^i{_\m})$. Using the field
equations to eliminate $\f^i{_\m}$, one can verify that $\cL_K$ reduces
to the standard BHT form.

Introducing the notation $Q_A=(b^i{_\m},\om^i{_\m},f^i{_\m}\l^i{_\m})$,
we now consider the linearized form of the theory  around a maximally
symmetric solution $\bar Q_A$, characterized by (Appendix A)
\bsubeq\lab{2.2}
\be
\barG_{ij}=\Leff\eta_{ij}\, ,\qquad \barf^i{_\m}=-\Leff\barb^i{_\m}\,,
\qquad \barl^i{_\m}=0\, ,                                  \lab{2.2a}
\ee
where $\Leff$ is the effective cosmological constant. The linearization
of the Lagrangian density \eq{2.1} is based on the expansion
\be
Q_A=\bar Q_A+\wt Q_A\,,
\ee
\esubeq
where $\wt Q_A$ is a small excitation around $\bar Q_A$. The piece
of $\cL$ quadratic in $\wt Q_A$ takes the form:
\bsubeq\lab{2.3}
\bea
\cL^{(2)}&=&a\ve^{\m\n\r}\left(2\s\tb^i{_\m}\bar\nab_\n\tom_{i\r}
  +\s\ve^{ijk}\bar b^i{_\m}\tom^j{_\n}\tom^k_{\r}
  -\L_0\ve_{ijk}\bar b^i{_\m}\tb^j{_\n}\tb^k{_\r}\right)   \nn\\
&&+\frac{a}{m^2}\cL_K^{(2)}
  +\ve^{\m\n\r}\tl^i{_\m}\left(\bar\nab_\n\tb_{i\r}
  +\ve_{ijk}\tom^j{_\n}\bar b^k{_\r}\right)\, ,
\eea
where
\bea
\cL_K^{(2)}&:=&\ve^{\m\n\r}\left(\tf^i{_\m}\bar\nab_\n\tom_{i\r}
  -\frac{\Leff}2 \ve_{ijk}\bar b^i{_\m}\tom^j{_\n}\tom^k{_\r}\right)
  -\left(b\cV_K\right)^{(2)}\, ,                           \nn\\
\left(b\cV_K\right)^{(2)}&:=&
  \frac{\bar b}{4}\left(\eta^{ij}\bar g^{\m\n}
  -\bar h^{i\m}\bar h^{j\n}\right)\tf_{i\m}\tf_{j\n}
  +\frac{\bar b}{2}\Leff\left(\eta^{ij}\bar g^{\m\n}
  +\bar h^{i\m}\bar h^{j\n}
  -2h^{i\n}h^{j\m}\right)\tf_{i\m}\tb_{j\n}                \nn\\
&&+\frac {\bar b}4\Leff^2\left(\eta^{ij}\bar g^{\m\n}-\bar
   h^{i\n}\bar h^{j\m}\right)\tb_{i\m}\tb_{j\n}\,.
\eea
\esubeq

\prg{Field equations.} The variation of $\cL^{(2)}$ with respect to
$\wt Q_A=(\tb^i{_\m},\tom^i{_\m},\tf^i{_\m},\tl^i{_\m})$ yields the
linearized BHT field equations:
\bea
&&a\ve^{\m\n\r}\left(2\s\bar\nab_\n\tom_{i\r}
  -2\bL\ve_{ijk}\bar b^j{}_\n\tb^k{}_\r\right)
  -\frac{a}{m^2}W_i{^\m}
  +\ve^{\m\n\r}\bar\nabla_\n\tl_{i\r}=0\, ,                \nn\\
&&\ve^{\m\n\r}\left[a\bar\nab_\n
  \left(2\s\tb_{i\r}+\frac{1}{m^2}\tf_{i\r}\right)
  +a\left(2\s-\frac{\Leff}{m^2}\right)\ve_{ijk}\bar b^j{_\n}\tom^k{_\r}
  +\ve_{ijk}\bar b^j{}_\n \tl^k{}_\r\right] =0\,,          \nn\\
&&\ve^{\m\n\r}\bar\nab_\n\tom_{i\r}
  -\frac {\bar b}2\left[(\eta_{ij}\bar g^{\m\n}
  -\bar h_i{^\m}\bar h_j{^\n})(\tf^j{_\n}+\Leff\tb^j{_\n})
  +2\Leff(\bar h_i{^\m}\bar h_j{^\n}
  -\bar h_i{^\n}\bar h_j{^\m})\tilde b^j{_\n}\right]=0\, , \nn\\
&&\ve^{\m\n\r}\left(\bar\nab_\n\tb_{i\r}
             +\ve_{ijk}\tom^j{_\n}\bar b^k{_\r}\right)=0\,.\lab{2.4}
\eea
where $W_i{^\m}:=\d\left(b\cV_K\right)^{(2)}/\d\tb^i{_\m}$ takes the
form:
\bea
W_i{^\m}&=&\frac{1}{2}\Leff\bar b
  \left[\left(\eta_{ij}\bar g^{\m\n}+\bar h_i{^\m}\bar h_j{^\n}
  -2\bar h_i{^\n}\bar h_j{^\m}\right)\tf^j{_\n}
  +\Leff(\eta_{ij}\bar g^{\m\n}
  -\bar h_i{^\n}\bar h_j{^\m})\tb^j{_\n}\right]\, .        \nn
\eea

Let us now focus our attention on the trace of the first field
equation, the linearized version of (A.3):
\be
\left(\s+\frac\Leff{2m^2}\right)
  \bar h_i{^\m}\left(\tf^i{_\m}+\Leff\tb^i{_\m}\right)=0\,.\lab{2.5}
\ee
In the canonical approach, this relation is expected to be a
\emph{constraint}, as is the case in the nonlinear regime. However,
there is a \emph{critical condition} on parameters, defined by
$\Leff+2\s m^2=0$, for which equation \eq{2.5} is identically
satisfied. This is an important signal that the related canonical
structure of the linearized theory might be significantly changed.
Using \eq{A.5}, the critical condition can be equivalently written as
\be
\L_0/m^2=-1\, ,                                            \lab{2.6}
\ee
or as $\Leff=2\s\L_0$. The central idea of our work is to examine the
influence of this condition on the \emph{canonical structure} of the
linearized BHT massive gravity.

\prg{Extra gauge symmetry.} When we have a maximally symmetric
background, the critical condition \eq{2.6} implies that the massive
graviton of the linearized BHT gravity (with two helicity states)
becomes a (single) partially massless mode; simultaneously, there
appears an extra gauge symmetry in the theory. By a systematic analysis
of the related canonical structure (see section 5), we discover that
this symmetry has the following form:
\bea
\d_E \tb^i{_\m}&=&\eps\barb^i{_\m}\, ,                     \nn\\
\d_E\tom^i{_\m}&=&
  -\ve^{ijk}\bar b_{j\m}\bar h_k{^\n}\bar\nab_\n\eps\,,    \nn\\
\d_E\tf^i{_\m}&=&-2\bar\nab_\m(\bar h^{i\n}\bar\nab_\n\eps)
                 +\Leff\eps\bar b^i{_\m}\,,                \nn\\
\d_E\tl^i{_\m}&=&0\, ,                                     \lab{2.7}
\eea
where $\eps$ is an infinitesimal gauge parameter. The proof of this
statement at the level of the field equations \eq{2.4} is given in
Appendix B. Although the form of $\d_E \tf^i{_\m}$ has been known for
some time, see for instance \cite{15,2}, our result uncovers the very
root of this symmetry by specifying its action on all the fields,
including $\tb^i{_\m}$. Up to second order terms, one can rewrite the
infinitesimal gauge transformation of $\tb^i{_\m}$ in the form $\d_E
b^i{_\m}=\eps b^i{_\m}$, which looks like a Weyl rescaling. However, in
doing so, one should keep in mind that \eq{2.7} is \emph{not} the
symmetry of the full nonlinear theory, but only of its linearized
version. Note also that the Weyl-like form of \eq{2.7} closely
resembles the result found in \cite{18}, which describes an extra
gauge symmetry of the Chern-Simons gravity. The presence of the gauge
parameter $\eps$ and its first and second derivatives in \eq{2.7}
indicates significant changes of the set of first class constraints, in
comparison to the nonlinear BHT theory.

\section{Canonical analysis of the linearized theory}
\setcounter{equation}{0}

We are now going to analyze the canonical structure of the BHT gravity
linearized around the maximally symmetric background
$G_{ij}=\Leff\eta_{ij}$, at the critical point \eq{2.6}. Technically,
the analysis is based on the Lagrangian \eq{2.3}, quadratic in the
excitation modes $\wt Q_A$.

\prg{Primary constraints.} If $P^A=\bar P^A+\wt P^A$ are the
canonical momenta conjugate to the field variables $Q_A=\bar Q_A+\wt
Q_A$, then transition to the linearized theory implies $\{\wt Q_A,\wt
P^B\}=\d_A^B$. In other words, the basic phase space variables of the
linearized theory are
$$
\wt Q_A=(\tb^i{_\m}, \tom^i{_\m},\tl^i{_\m},\tf^i{_\m})\, ,\qquad
\wt P^A=(\tpi_i{^\m},\wt\Pi_i{^\m},\tp_i{^\m},\tP_i{^\m})\, .
$$
From the Lagrangian \eq{2.3}, we obtain the primary constraints of the
linearized theory:
\bea
&&\phi_i{^0}:=\tilde\pi_i{^0}\approx 0\, ,\qquad\,\,
  \phi_i{^\a}:=\tilde\pi_i{^\a}
  -\ve^{0\a\b}\tilde\l_{i\b}\approx 0\, ,                  \nn\\
&&\Phi_i{^0}:=\wt\Pi_i{^0}\approx 0\, ,\qquad
  \Phi_i{^\a}:=\wt\Pi_i{^\a}-2a\ve^{0\a\b}
  \left(\s\tilde b_{i\b}+\frac{1}{2m^2}\tilde\f_{i\b}\right)
  \approx 0\, ,                                            \nn\\
&&\tp_i{^\m}\approx 0\, ,\hspace{61pt}
  \tP_i{^\m}\approx 0\, .                                  \lab{3.1}
\eea

\prg{Total Hamiltonian.} Inspired by the results of~\cite{11}, we find
that the quadratic canonical Hamiltonian $\cH_c$  can be represented in
the form (up to a divergence):
\bea
\cH_c&=&\tb^i{_0}\cH_i+\tom^i{_0}\cK_i+\tf^i{_0}\cR_i+\tl^i{_0}\cT_i\nn\\
  &&+\bar b^i{_0}{\cal A}_i+\bar\om^i_0{\cal B}_i
    +\bar f^i{_0}{\cal C}_i+\frac{a}{m^2}(b\cV_K)^{(2)}\,. \lab{3.2}
\eea
The components of $\cH_c$ are defined as follows:
\bea
&&\cH_i:=-\ve^{0\a\b}\left(2a\s\bar\nab_\a\tom_{i\b}
  -2a\bL\ve_{ijk}\bar b^j{}_\a \tb^k{}_\b
  +\bar\nabla_\a\tl_{i\b}\right)\, ,\,,                    \nn\\
&&\cK_i:=-\ve^{0\a\b}\left[a\bar\nab_\a
           \left(2\s \tb_{i\b}+\frac{1}{m^2}\tf_{i\b}\right)
  +a\left(2\s-\frac{\Leff}{m^2}\right)\ve_{ijk}\bar b^j{_\a}
   \tom^k{_\b}+\ve_{ijk}\bar b^j{}_\a\tl^k{}_\b\right]\, , \nn\\
&&\cR_i:=-\frac{a}{m^2}\ve^{0\a\b}\bar\nab_\a\tom_{i\b}\, ,\nn\\
&&\cT_i:=-\ve^{0\a\b}\left(\bar\nab_\a\tb_{i\b}
  +\ve_{ijk}\tom^j{_\a}\bar b^k{_\b}\right)\, ,            \nn\\
&&{\cal A}_i:=-\ve^{0\a\b}\ve_{ijk}
  \left(a\s\tom^j{_\a}\tom^k{_\b}
  -a\L_0\tb^j{_\a}\tb^k{_\b}+\tom^j{_\a}\tl^k{_\b}\right)\,,\nn\\
&&{\cal B}_i:=-\ve^{0\a\b}\ve_{ijk}\left(2a\s \tb^j{_\a}\tom^k{_\b}
  +\frac{a}{m^2}\tom^j{_\a}\tf^k{_\b}+\tb^j{_\a}\tl^k{_\b}\right)\,,\nn\\
&&{\cal C}_i:=-\frac{a}{2m^2}\ve^{0\a\b}\ve_{ijk}\tom^j{_\a}\tom^k{_\b}\,.
\eea

In order to simplify further exposition, we find it more convenient to
continue our analysis in a reduced phase space formalism. The formalism
is based on using the 24 second class constraints
$X_A=(\phi_i{^\a},\Phi_i{^\a},\tp_i{^\a},\tP_i{^\a})$ to eliminate the
momenta $(\tpi_i{^\a},\wt\Pi_i{^\a},\tp_i{^\a},\tP_i{^\a})$. The
dimension of the resulting reduced phase space $R_1$ is $N=72-24=48$,
and its structure is defined by the basic nontrivial Dirac brackets
(DB):
\bea
&&\{\tb^i{_\a},\tl^j{_\b}\}^*_1=\eta^{ij}\ve_{0\a\b}\d\, ,\qquad
 \{\tom^i{_\a},\tf^j{_\b}\}^*_1=\frac{m^2}{a}\eta^{ij}\ve_{0\a\b}\d\,,\nn\\
&&\{\tl^i{_\a},\tf^j{_\b}\}^*_1=-2m^2\s\eta^{ij}\ve_{0\a\b}\d\,,
\eea
while the remaining DBs remain the same as the corresponding Poisson
brackets. In $R_1$, the total Hamiltonian takes the form:
\be
\cH_T=\cH_c+u^i{_0}\phi_i{^0}+v^i{_0}\Phi_i{^0}
           +w^i{_0}\tp_i{^0}+z^i{_0}\wt P_i{^0}\, .        \lab{3.4}
\ee

\prg{Secondary constraints.} The consistency conditions of the primary
constraints $\tpi_i{^0},\tPi_i{^0},\tp_i{^0}$ and $\tP_i{^0}$ produce
the secondary constraints:
\bsubeq\lab{3.6}
\bea
\hcH_i&:=&\cH_i+\frac{a}{m^2}W_i{^0}\approx 0\, ,          \nn\\
\cK_i&\approx& 0 \, ,                                      \lab{3.6a}\\
\hcR_i&:=&\cR_i+\frac{a\bar b}{2m^2}
  \left[(\eta_{ij}\bar g^{0\m}-\bar h_i{^0}\bar h_j{^\m})(\tf^j{_\m}
  +\Leff\tb^j{_\m})\right.                                 \nn\\
&&\hspace{60pt}
  +\left.2\Leff(\bar h_i{^0}\bar h_j{^\m}
  -\bar h_i{^\m}\bar h_j{^0})\tilde b^j{_\m}\right]\approx 0\, ,\nn\\
\cT_i&\approx& 0 \, .                                      \lab{3.6b}
\eea
\esubeq

\prg{Tertiary constraints.} Let us now introduce the change of
variables:
\be
z^i{_0}'=z^i{_0}-\bar f^i{_m}u^m{_0}\, ,\qquad
\tilde\p_i{^0}'=\tilde\p_i{^0}+\bar f_i{^k}\tP_k{^0}\, ,
\ee
such that
$$
u^i{_0}\tilde\p_i{^0}+z^i{_0}\tP_i{^0}
  =u^i{_0}\tilde\p{}_i{^0}'+z^i{_0}'\tP_i{^0}\, .
$$
The consistency conditions of $\cK_i$ and $\hcR_i$ determine two
components $z'_{\b 0}:=\bar b^k{_\b}z'_{k0}$ of $z'_{k0}$:
\bea
z'_{\b0}=-\ve_{ijk}\bar b^i{_0}\bar\om^j{_0}(\tf^k{_\b}
  +\Leff\tb^k{_\b})+\bar b_{i0}\bar\nab_\b(\tf^i{_0}+\Leff \tb^i{_0})
  +\frac{m^2}a\ve_{ijk}\bar b^i{_0}\bar b^j{_\b}\tl^k{_0}\,,\nn
\eea
while the consistency of $\hcH_i$ and $\cT_i$ leads to the tertiary
constraints:
\bsubeq
\bea
&&\th_{\m\n}:=\tf_{\m\n}-\tf_{\n\m}\approx 0\, ,           \lab{3.8a}\\
&&\psi_{\m\n}:=\tl_{\m\n}-\tl_{\n\m}\approx 0\, ,          \lab{3.8b}
\eea
where
\be
\tf_{\m\n}= \bar b^i{_\m}\tf_{i\n}-\Leff\bar b^i{_\n}\tb_{i\m}\,,
\qquad \tl_{\m\n}=\bar b^i{_\m}\tl_{i\n}\,.
\ee
\esubeq

\prg{Quartic constraints.} Further consistency conditions determine two
components $w_{\b 0}:=\bar b^k{_\b}w_{k0}$ of $w_{k0}$:
\bea
w_{\b0}&=&-\ve_{ijk}\bar b^i{_0}\bar\om^j{_0}\tl^k{_\b}
  +\bar b^i{_0}\bar\nab_\b \tl^i{_0}
  -2\L_0\ve_{ijk}\bar b^i{_0}\bar b^j{_\b}\tb^k{_0}
  +\frac a{m^2}\ve_{0\b\a}\bar b^i{_0}W_i{^\a}             \nn\\
&&-a\s\bar b\ve_{0\b\a}\left(\bar b_{i0}\bar g^{\a\n}(\tf^i{_\n}
  +\Leff\tb^j{_\n})-2\Leff\bar h_i{^\a}\tb^i{_0}\right)\,, \nn
\eea
and produce the relations
\bsubeq
\bea
&&\chi:=\bar h_i{^\m}\tl^i{_\m}\approx 0\, ,         \lab{3.9a}\\
&&\vphi:=\left(\s+\frac\Leff{2m^2}\right)
  \bar h_i{^\m}\left(\tf^i{_\m}+\Leff\tb^i{_\m}\right)\approx 0\, .
\eea
\esubeq
At the critical point \eq{2.6}, the expression $\vphi$ identically
vanishes, and the only quartic constraint is $\chi$.

We close the consistency procedure by noting that the consistency of
$\chi$ determines the multiplier $w_{00}:=\bar b^k{_0}w_{k0}$:
$$
\bar g^{00}w_{00}=-\left(2\bar g^{\a 0}w_{\a0}
+\bar g^{\a\b}\bar b^i{_\a}\dot\tl_{i\b}\right)\,,
$$
where $\dot\tl_{i\b}$ is calculated in Appendix C, while the absence of
$\vphi$ implies that $z'_{00}:=\bar b^k{_0}z'_{k0}$ remains
undetermined.

\bitem
\item[$\bull$] In comparison to the nonlinear BHT massive gravity,
the linearized theory has \emph{one constraint less} ($\vphi$) and
\emph{one undetermined multiplier more} ($z'_{00}$), which leads to a
significant modification of its canonical structure.
\eitem

\section{Classification of constraints}
\setcounter{equation}{0}

Among the primary constraints, those that appear in $\cH_T$ with
arbitrary multipliers ($u^i{_0},v^i{_0}$ and $z'_{00}$) are first
class (FC):
\be
\tpi_i{^0}{}',\wt\Pi_i{^0}, \tP^{00}=\mbox{FC}\, ,
\ee
while the remaining ones, $\tilde p_i{^0}$ and $\tP^{\a0}$, are second
class. Note that $\tP^{00}:=\bar h^{k0}\tP_k{^0}$.

Going to the secondary constraints, we use the following simple
theorem:
\bitem
\item[$\bull$] If $\phi$ is a FC constraint, then $\{\phi,H_T\}^*_1$
is also a FC constraint.
\eitem
The proof relies on using the Jacoby identity. The theorem implies that
the secondary constraints $\hcH'_i:=-\{\tpi_i{^0}{}',H_T\}^*_1$,
$\cK_i=-\{\wt\Pi_i{^0},H_T\}^*_1$ and
$\hcR^{00}:=-\{\tP^{00},H_T\}^*_1$ are FC. A straightforward
calculation yields:
\bea
\hcH_i'&=&\hcH_i+\bar f_i{^k}\hcR_k\, ,                    \nn\\
\hcR^{00}&=&\bar h_i{^0}\hcR{}^i
 -\bar h_i{^0}\bar\nab_\b(\bar b^i{_0}\tP^{\b0})
 +\frac{a}{2m^2}\bar b\ve_{0\a\b}
  \frac{\bar g^{0\a}}{\bar g^{00}}(\bar f^{0\b}
 -\bar g^{00}\bar f_0{^\b})\tilde p^{00}                   \nn\\
 &&-a\bar b\ve_{0\a\b}\left[\s \bar g^{0\b}
  +\frac{1}{2m^2}(\bar f^{0\b}-\bar g^{0\b}\bar f
  +\bar g^{0\b}\bar f_0{^0}-\bar g^{00}\bar f_0{^\b})
   \right]\tp^{\a0}\, .                                    \nn
\eea
Since the background is maximally symmetric, we have:
\bea
&&\hcH'_i=\hcH_i-\Leff\hcR_i\, ,                           \nn\\
&&\hcR^{00}=\bar h_i{^0}\hcR{}^i
  -\bar h_i{^0}\bar\nab_\b(\bar b^i{_0}\tP^{\b0})\, .
\eea
After identifying the above 14 FC constraints, we now turn our
attention to the remaining (tertiary and quartic) 17 constraints.
However, we know \cite{10} that the number of second class constraints
has to be even. As one can verify, the constraint $\wt\psi_{\a\b}$ is
FC, while the other 16 constraints are second class (Appendix D). The
complete classification of constraints in the reduced space $R_1$ is
displayed in Table 1.
\begin{center}
\doublerulesep 1.8pt
\begin{tabular}{lll}
\multicolumn{3}{l}{\hspace{16pt}Table 1. Classification
                                         of constraints in $R_1$} \\
                                                      \hline\hline
\rule{0pt}{12pt}
&~First class \phantom{x}&~Second class \phantom{x} \\
                                                      \hline
\rule[-1pt]{0pt}{16pt}
\phantom{x}Primary &~$\tpi_i{^0}'',\tPi_i{^0},\tP^{00}$
            &~$\tp_i{^0},\tP^{\a0}$        \\
                                                      \hline
\rule[-1pt]{0pt}{19pt}
\phantom{x}Secondary\phantom{x} &~$\hcH'_i,\cK_i,\hcR^{00}$
           &~$\cT_i,\hcR{}^\a{}'$             \\
                                                      \hline
\rule[-1pt]{0pt}{16pt}
\phantom{x}Tertiary\phantom{x}
   &~ $\psi_{\a\b}$&~$\th_{0\a},\th_{\a\b},\psi_{0\a}$ \\
                                                      \hline

\rule[-1pt]{0pt}{16pt}
  \phantom{x}Quartic\phantom{x}
  & &~$\chi$\\                                  \hline\hline
\end{tabular}
\end{center}
Here, $\hcR^\a{}'=\bar h_i{^\a}\hcR^i{}'$, where $\hcR_i{}'$ is a
suitable modification of $\hcR_i$, defined so that it does not contain
$\tf_{i0}$:
\bea
\hcR_i{}'&:=& \hcR_i{}-\frac{a\bar b}{4m^2}\left(
  \bar h_i{^\m}\bar g^{0\n}-\bar h_i{^\n}\bar g^{0\m}\right)\th_{\m\n}\nn\\
  &\equiv&\cR_i+\frac{a\bar b}{2m^2}(\bar h_i{^\a}\bar h_j{^0}
  -\bar h_i{^0}\bar h_j{^\a})(\tf^j{_\a}-\Leff \tb^j{_\a})\,.
\eea

Now, we can calculate the number of independent dynamical degrees of
freedom with the help of the standard formula:
$$
N^* = N-2N_1-N_2\, ,
$$
where $N$ is the number of phase space variables in $R_1$, $N_1$ is the
number of FC, and $N_2$ the number of second class constraints. Using
$N=48$ and, according to the results in Table 1, $N_1 = 15$ and $N_2 =
16$, we obtain that
\bitem
\item[$\bull$] the number of physical modes in the phase space is
$N^* = 2$, and consequently, the BHT theory at the critical point
\eq{2.6} exhibits one Lagrangian degree of freedom.
\eitem

\section{Extra gauge symmetry}
\setcounter{equation}{0}

The presence of an extra primary FC constraint $\wt P^{00}$ implies the
existence of an extra gauge symmetry. To simplify its canonical
construction, we go over to the reduced phase space $R_2$, which is
obtained from $R_1$ by using the additional constraints
\be
R_2:\qquad\th_{\b 0}\equiv \tf_{\b 0}-\tf_{0\b}=0\, ,
\qquad \wt P^{\b 0}=0\, ,
\ee
to eliminate the variables $\tf_{\b0}$ and $\wt P^{\b0}$. Basic DBs
between the canonical variables in $R_2$ retain the same form as in
$R_1$. Starting with the primary FC constraint $\wt P^{00}$,
Castellani's algorithm \cite{19} leads to the following canonical
generator in $R_2$:
\bea
G_E&=&-2\ddot{\eps}\wt P^{00}
  +\dot\eps\left[-2\hcR^{0}{}'
  +2(\bar h^{i0}\bar \nab_0 \bar b^i{_0})\tP^{00}
  +\ve_{ijk}\bar h^{i0}\bar b^j{_0}\wt\Pi^k{_0}\right]     \nn\\
&&+\eps\left[\ve^{0\a\b}\bar b^i{_\a}\tl_{i\b}+\tpi_0{^0}
  -\ve_{ijk}\bar\nab_\a(\bar h^{i\a}\bar b^j{_0}\wt\Pi^k{_0})
  +2\bar\nab_\a\hcR^\a{}'\right.                      \nn\\
&&\qquad\left.
  -2\bar\nab_\a(\bar h_i{^\a}\tP^{00}\bar\nab_0\bar b^i{_0})
  +\Leff \bar g_{00}\tP^{00}\right]\,.
\eea

The action of $G_E$ on $R_2$ is given by $\d_E\phi=\{\phi,G_E\}_2^*$,
which yields:
\bsubeq\lab{5.4}
\bea
\d_E \tb^i{_\m}&=&\eps\bar b^i{_\m}\, ,                    \nn\\
\d_E\tom^i{_\m}
  &=&-\ve^{ijk}\bar b_{j\m}\bar h_k{^\n}\bar\nab_\n\eps\,, \nn\\
\d_E\tf^i{_\a}&=&-2\bar\nab_\a(\bar h^{i\n}\bar\nab_\n\eps)
  +\Leff\eps \bar b^i{_\a}\, ,                             \nn\\
\d_E\tf_{00}&=&
  -2b_{i0}\bar\nab_0\left(\bar h^{i\n}\bar\nab_\n\eps\right)\,,\nn\\
\d_E\tl^i{_\m}&=&0\,.
\eea

To make a comparison with \eq{2.5}, we now derive the transformation
law for the variable
$\tf^i{_0}=\bar h_i{^\m}\tf_{\m0}
  +\Leff\bar h_i{^\n}\bar b^j{_0}\tb_{j\n}$.
Using
\bea
\d_E\tf_{\b0}&=&-2\bar b^i{_0}\bar\nab_\b(\bar h_i{^\n}\bar\nab_\n\eps)\nn\\
&=&-2\pd_\b\pd_0\eps-2\bar b^i{_0}(\bar\nab_\b \bar h_i{^\n})\pd_\n\eps\nn\\
&=&-2\pd_0\pd_\b\eps-2\bar b^i{_\b}(\bar\nab_0 \bar h_i{^\n})\pd_\n\eps\nn\\
&=&-2\bar b^i{_\b}\bar\nab_0(\bar h_i{^\n}\bar\nab_\n\eps)\,,\nn
\eea
one obtains
\be
\d_E\tf^i{_0}=-2\bar\nab_0(\bar h^{i\n}\bar\nab_\n\eps)
  +\Leff\eps\bar b^i{_0}\,.
\ee
\esubeq
The transformation rules \eq{5.4} are in complete agreement with
\eq{2.5}.

\section{Concluding remarks}
\setcounter{equation}{0}

In the \emph{nonperturbtive} regime of the BHT gravity, the constraint
structure is found to depend critically on the value of $\Om^{00}$,
where $\Om^{\m\n}=\s g^{\m\n}+G^{\m\n}/2m^2$ \cite{11}. In the region
of the phase space where $\Om^{00}\ne 0$, the BHT theory has \emph{two}
Lagrangian degrees of freedom, which corresponds to two helicity states
of the massive graviton excitation.

In this paper, we studied the canonical structure of the BHT gravity
\emph{linearized} around the maximally symmetric background,
$G^{\m\n}=\Leff g^{\m\n}$. At the critical point $\L_0/m^2=-1$, the
background solution is characterized by the property $\Om^{\m\n}=0$,
the covariant version of $\Om^{00}=0$. Analyzing the constraint
structure of the linearized theory, we constructed the canonical
generator of \emph{extra gauge symmetry}, which is responsible for
transforming two massive graviton excitations into a single, partially
massless mode; moreover, the theory is found to have \emph{one}
Lagrangian degree of freedom.

In order to properly understand the linearized theory, one should
stress that although we have $\Om^{\m\n}=0$ on the very background, the
linearized theory is well-defined in the region off the background,
where $\Om^{\m\n}\ne 0$. In this region, the process of linearization
induces a drastic modification of the canonical structure of the BHT
theory, leading to the change of the number and type of constraints and
physical degrees of freedom.

Thus, the canonical structure of the BHT gravity at the critical point
$\L_0/m^2=-1$ does not remain the same after linearization. Following
the arguments of Chen et al. \cite{20}, we are led to conclude that
the canonical consistency of the BHT gravity, expressed by the
stability of its canonical structure under linearization, is violated
at the critical point $\L_0/m^2=-1$.

\section*{Acknowledgements}

This work was supported by the Serbian Science Foundation under Grant
No. 171031.

\appendix
\section{Maximally symmetric solutions}
\setcounter{equation}{0}

Variation of  the action \eq{2.1} with respect to the basic dynamical
variables $b^i{_\m},\om^i{_\m},f^i{_\m}$ and $\l^i{_\m}$, yields the
following field equations \cite{11}:
\bea
&&a\ve^{\m\n\r}\left(\s R_{i\n\r}-\bL\ve_{ijk}b^j{_\n}b^k{_\r}\right)
  -\frac{ab}{m^2}\cT_i{^\m}+\ve^{\m\n\r}\nab_\n\l_{i\r}=0\,,\nn\\
&&\ve^{\m\n\r}\left(aT_{i\n\r}+\frac{a}{m^2}\nab_\n f_{i\r}
  +\ve_{ijk}b^j{_\n}\l^k{_\r}\right)=0\,,                   \nn\\
&&\ve^{\m\n\r}R_{i\n\r}-b\left(f_i{^\m}-f h_i{^\m}\right)=0\,,\nn\\
&&\ve^{\m\n\r}T_{i\n\r}=0\,,                               \lab{A.1}
\eea
where $\cT_i{^\m}$ is the energy-momentum tensor associated to $\cL_K$,
$$
\cT_i{^\m}:=-\frac{1}{b}\frac{\pd\cL_K}{\pd b^i{_\m}}
           =h_i{^\m}\cV_K-\frac{1}{2}(f_{ik}f^{k\m}-ff_i{^\m})\, .
$$
The last equation ensures that spacetime is Riemannian, the third and
the second one imply
$$
f_{ij}=2L_{ij}\, ,\qquad \l_{ij}=\frac{2a}{m^2}C_{ij}\,,
$$
where  $L_{ij}$ and $C_{ij}$ are the Schouten and the Cotton tensor,
respectively:
$$
L_{ij}=\hR_{ij}-\frac{1}{4}\eta_{ij}R\,,\qquad
C_{ij}=\ve_i{}^{mn}\nabla_m L_{nj}\, ,
$$
and the first field equation takes the form:
\be
\s G_{ij}-\bL\eta_{ij}-\frac{1}{2m^2}K_{ij}=0\, ,        \lab{A.2}
\ee
where $K_{ij}:=\cT_{ij}-2(\nabla_m C_{in})\ve^{mn}{_j}$. We display
here also a set of algebraic consequences of the field equations:
\bea
&&\f_{\m\n}=\f_{\n\m}\, ,                                  \nn\\
&&\l_{\m\n}=\l_{\n\m}\, ,\qquad \l=0\, ,                   \nn\\
&&\s\f+3\bL+\frac{1}{2m^2}\cV_K=0\, .                      \lab{A.3}
\eea

For maximally symmetric solutions with $\bar G_{ij}=\Leff\eta_{ij}$,
we have $\bar L_{ij}=-\frac 12\Leff\eta_{ij}$, $\bar C_{ij}=0$, and
consequently:
\be
\bar f_{ij}=-\Leff\eta_{ij}\,,\qquad \bar\l_{ij}=0\,.      \lab{A.4}
\ee
Then, the last equation in \eq{A.3} gives the following solution for
$\Leff$:
\be
\Leff=-2m^2\left(\s\pm\sqrt{1+\bL/m^2}\right)\, .          \lab{A.5}
\ee
For $\L_0/m^2=-1$, the ground state is uniquely defined by
$\Leff=-2m^2\s$.

\section{Lagrangian form of extra gauge symmetry}
\setcounter{equation}{0}

In this appendix, we show that the linearized field equations \eq{2.4}
are invariant under the extra gauge symmetry \eq{2.7}. The last two
equations in \eq{2.4} are invariant for all values of the parameters.
Denoting the left-hand sides of the first two field equations by $F_1$
and $F_2$, respectively, we find:
\bea
&&\d_E F_1=-4a\left(\bL+\frac{\Leff^2}{4m^2}\right)
                 \bar b \bar h_i{^\m}\eps
   +a\left(2\s+\frac{\Leff}{m^2}\right)
   \ve^{\m\n\r}\ve_{ijk}\bar b^j{_\n}\bar\nab_\r
   (\bar h^{k\s}\bar\nab_\s\eps)\, ,                       \nn\\
&&\d_E F_2=-2a\left(2\s+\frac{\Leff}{m^2}\right)
   \ve^{\m\n\r}\bar b_{i\n}\bar\nab_\r\eps\, .             \nn
\eea
The corresponding conditions of invariance,
$$
2\s+\frac{\Leff}{m^2}=0\,,\qquad \bL+\frac{\Leff^2}{4m^2}=0\,,
$$
are both equivalent to the critical condition \eq{2.6}.

\section{Calculation of the determined multipliers}
\setcounter{equation}{0}

In the process of calculating the values of the multipliers $z'_{\b
0},w_{\b 0}$ and $w_{00}$, we need the DBs of $\wt
Q_A=(\tb^i{_\a},\tom^i{_\a},\tl^i{_\a},\tf^i{_\a})$ with the total
Hamiltonian:
\bea
\dot \tb^i{_\a}
  &=&-\ve^{ijk}(\bar\om_{j0}\tb_{k\a}+\tilde \om_{j0}\bar b_{k\a}
    -\tom_{j\a}\bar b_{k0})+\bar\nab_\a \tb^i{_0}\, ,      \nn\\
\dot\tom^i{_\a}&=&-\ve^{ijk}\bar\om_{j0}\tom_{k\a}
  +\bar\nab_\a\tom^i{_0}                                   \nn\\
 &&+\frac{1}{2}\bar b\ve_{0\a\b}\left[(\d^i{_j}\bar g^{\b\n}
  -\bar h^{i\b}\bar h_j{^\n})(\tf^j{_\n}+\Leff\tb^j{_\n})
  +2\Leff(\bar h^{i\b}\bar h_j{^\n}
    -\bar h^{i\n}\bar h_j{^\b})\tilde b^j{_\n}\right]\, ,  \nn\\
\dot\tl^i{_\a}&=&-\ve^{ijk}\bar\om_{j0}\tl_{k\a}
  +\bar\nab_\a\tl^i{_0} +2\L_0\ve^{ijk}(\bar b_{j0}\tb_{k\a}
  -\bar b_{j\a}\tb_{k0})+\frac{a}{m^2}\ve_{0\a\b}\tilde W_i{^\b}\nn\\
&&-a\s \bar b\ve_{0\a\b}\left[(\d^i{_j}\bar g^{\b\n}
  -\bar h^{i\b}\bar h_j{^\n})(\tf^j{_\n}+\Leff\tb^j{_\n})
  +2\Leff(\bar h^{i\b}\bar h_j{^\n}
    -\bar h^{i\n}\bar h_j{^\b})\tilde b^j{_\n}\right]\, ,  \nn\\
\dot\tf^i{_\a}&=&-\ve^{ijk}(\bar\om_{j0}\tf_{k\a}
  +\tilde\om_{j0}\bar f_{k\a}-\tom_{j\a}\bar f_{k0})
  +\bar\nab_{\a}\tf^i{_0}-\frac{m^2}{a}\ve^{ijk}(\bar b_{j0}\tl_{k\a}
  -\bar b_{j\a}\tl_{k0})\, .                               \nn
\eea

\section{Second class constraints}
\setcounter{equation}{0}

In this appendix, we show that the set of 16 constraints in the second
column of Table 1 is of the second class. Instead of calculating the
determinant of the $16\times 16$ matrix of the related DBs, the proof
is derived iteratively.

\prg{Step 1. }We begin with the subset of constraints
$Y_A:=(\psi_{\b0},\chi,\tp^{\a 0}, \tp_0{^0})$. The corresponding $6\times
6$ matrix $\D_1$ with matrix elements $\{Y_A,Y_B\}^*_1$ reads:
$$
\D_1=\left(\ba{cc}
           0_{3\times 3} & -I_{3\times 3} \\[5pt]
          -I^T_{3\times 3} & 0_{3\times 3}
          \ea\right)\, ,
$$
where $I$ is the unit matrix. The matrix $\D_1$ is regular, $\det\D_1=1$.

\prg{Step 2.} Next, we consider the subset of constraints
$Z_A:=(\th_{0\b},P^{\a 0})$. The corresponding $4\times 4$ matrix
\be
\D_2=\left(\ba{cc}
             0_{2\times 2} & \d_\b^\a \\[5pt]
            -\d_\b^\a & 0_{2\times 2}
             \ea\right)                                     \nn
\ee
is regular, since $\det\D_2=1$.

\prg{Step 3.} Finally, we consider the remaining subset
$W_A=(\cT_i,\hcR^\a{}',\frac{1}{2}\ve^{0\a\b}\th_{\a\b})$. The $6\times
6$ matrix $\{W_A,W_B\}^*_1$ takes the form
$$
\D_3=\left(\ba{ccc}
           0_{3\times 3} & D_{3\times 2}& E_{3\times 1} \\
          -D^T_{2\times 3} & F_{2\times 2}&0_{2\times 1}\\
           -E^T_{1\times 3}&0_{1\times 2}&0_{1\times 1}\ea\right)\, ,
$$
where the matrices $D,E$ and $F$ are given by
\bea
D_i{^\a}&:=&\{\cT_i,\hcR^\a{}'\}^*_1
  =-\bar b \bar h^{j\a}\left[\ve_{ijn}\left(
    \frac{1}{2}\bar h^{n0}-\bar g^{00}\bar b^n{_0}\right)
  -\bar h_j{^0}\ve_{imn}\bar b^m{_0}\bar h^{n0}\right]\d\,,\nn\\
E_i&:=&\{\cT_i,\frac{1}{2}\ve^{0\a\b}\th_{\a\b}\}^*_1
      =-\frac{2m^2}{a}\bar b\bar h_i{^0}\d\, ,             \nn\\
F^{\a\b}&:=&\{\hcR^\a{}',\hcR^\b{}'\}_1^*\, .    \nn
\eea
The regularity of $\D_3$ follows from
\bea
&&\det\D_3=\left(
   \frac{1}{2}\ve^{ijk}\ve_{0\a\b}E_i D_j{^\a}D_k{^\b}\right)^2
  =\left(\frac{m^2}{2a}\bar b^2\right)^2\neq 0\, .                                  \nn
\eea

\end{document}